\documentclass{article}

    \usepackage[final]{neurips_2024}

\usepackage[utf8]{inputenc} %
\usepackage[T1]{fontenc}    %
\usepackage{hyperref}       %
\usepackage{url}            %
\usepackage{booktabs}       %
\usepackage{amsfonts}       %
\usepackage{nicefrac}       %
\usepackage{microtype}      %
\usepackage{xcolor}         %

\usepackage{graphicx}
\usepackage{svg}
\usepackage{subcaption}

\makeatletter
\newcommand*{\rom}[1]{\expandafter\@slowromancap\romannumeral #1@}
\makeatother

\usepackage{enumitem}
\setlist[itemize]{leftmargin=*,labelindent=0.0em,labelsep=-0.5cm}

\usepackage[autostyle]{csquotes}
\usepackage{setspace}       

\SetBlockEnvironment{innerquote}
\renewcommand{\mkbegdispquote}[2]{\textooquote}

\usepackage{blindtext} %
\usepackage{epigraph} 

\usepackage{listings}
\title{Charting the Shapes of Stories with Game Theory}

\author{%
    Constantinos Daskalakis \\
    MIT CSAIL \& Archimedes AI\\
    Cambridge, MA \\
    \texttt{costis@csail.mit.edu} \\
    \And
    Ian Gemp\\
    Google DeepMind \\
    London, UK \\
    \texttt{imgemp@google.com} \\
    \And
    Yanchen Jiang \\
    Harvard SEAS \\
    Cambridge, MA \\
    \texttt{yanchen\_jiang@g.harvard.edu} \\
    \And
    Renato Paes Leme \\
    Google Research \\
    New York, New York \\
    \texttt{renatoppl@google.com} \\
    \And
    Christos Papadimitriou \\
    Columbia CS \& Archimedes AI \\
    New York, NY \\
    \texttt{christos@columbia.edu} \\
    \And
    Georgios Piliouras \\
    Google DeepMind \\
    London, UK \\
    \texttt{gpil@google.com} \\
}

\begin{document}

\maketitle

\begin{abstract}
Stories are records of our experiences and their analysis reveals insights into the nature of being human. Successful analyses are often interdisciplinary, leveraging mathematical tools to extract structure from stories and insights from structure. Historically, these tools have been restricted to one dimensional charts and dynamic social networks; however, modern AI offers the possibility of identifying more fully the plot structure, character incentives, and, importantly, counterfactual plot lines that the story could have taken but did not take. In this work, we use AI to model the structure of stories as game-theoretic objects, amenable to quantitative analysis. This allows us to not only interrogate each character's decision making, but also possibly peer into the original author's conception of the characters' world. We demonstrate our proposed technique on Shakespeare's famous Romeo and Juliet. We conclude with a discussion of how our analysis could be replicated in broader contexts, including real-life scenarios.
\end{abstract}

\section{Introduction}

Stories have long been one of humanity's most cherished methods for transmitting knowledge, taking shape in epic poems, religious texts, folktales, literature, and the oral traditions of cultures, families, and organizations. While stories are remarkably effective as a communication tool between humans, they were, until very recently, challenging for machine learning models to process due to their lack of a clear structure that facilitates mathematical reasoning.

There have been various attempts to extract from stories a mathematical structure, and then use that to reason quantitatively about the story.  Kurt~\cite{vonnegut1999palm} proposed in his avant-garde (and originally rejected) master’s thesis the notion of the shapes of a story. In his famous lecture, he says:

\begin{displayquote}
I have tried to bring scientific thinking to literary criticism... Stories have very simple shapes, ones that computers can understand.
\end{displayquote}

Vonnegut’s attempt consisted in constructing one dimensional functions that represented the rise and fall of characters’ fortunes across a plot line and then classifying stories according to the shapes of such functions. More sophisticated methods have later tried to represent the relationships between the characters as a dynamic graph (see the survey by~\cite{labatut2019extraction}) and used tools from network analysis to dissect stories according to the graphs they induce. \cite{reagan2016emotional} constructs shapes by analyzing how the overall sentiment varies across different windows of text within the story. \cite{toubia2021quantifying} again splits the story in different text windows and studies the geometry of the set of points obtained by the respective embeddings, obtaining notions of speed, volume and circuitousness. \cite{boyd2020narrative} extracts a narrative structure by tracking the rate in which certain categories of words appear along the story arc.

In this paper we make a similar attempt to extract a mathematical structure, except that this time we focus on the main {\em decisions} that the characters face, and the incentives and beliefs driving their actions. To that end, we will represent stories as (extensive-form) games and then apply tools from game theory such as equilibrium computation to understand stories.

\paragraph{Stories as Games:} We will view stories as a sequence of decisions by different characters that affect their ultimate fate. We get to learn what each character decided and how the events unfolded. It is usually left to the reader to imagine what would have happened in the alternative scenarios (what if Romeo and Juliet were to live happily together?). It is also left to the imagination of the reader what was the motivation of the characters to act how they did (how would Juliet feel if she married Paris instead of Romeo?). Those are in fact closely related, as decisions can be seen as expressions of how characters would evaluate counterfactual scenarios.

In this paper, we present a proof of concept on how to use large language models (LLMs) to reconstruct this hidden structure that encodes both the counterfactual scenarios as well as the incentives of the characters to act the way they did. We use LLMs to extract the main characters in a story, the main decisions they face and their payoffs in alternative plot lines. Using those, we are able to represent them formally as an extensive-form game. The story itself will correspond to one of the sample paths of the game tree. We will then use a game solver to compute an equilibrium of the induced game and relate it to the story. We will say that a game ``rationalizes'' a story if the sample path corresponding to the actual story occurs with positive probability in equilibrium.

This is of course a highly ambiguous task. Different readers will interpret Juliet's motives differently. We view this as a feature, not a bug. In this preliminary version, we will demonstrate the steps to produce one way of rationalizing a story. However, our methodology allows us to recover multiple game trees that can rationalize the story in different ways.

For the purposes of illustration we choose a story that is widely known (namely, Romeo and Juliet) where the decisions are taken by fictional characters. The methodology showcased here can be equally applied if instead of a literary text we had a business email and we are analyzing the decisions faced by the executive of an organization or a salesperson.

\paragraph{Analyzing the Shape of Stories / Suspense and Surprise:} Once a story is represented as a game, we can use the payoffs and the equilibrium probabilities to determine the expected fortune of each character along the story as well as the likelihood of the sampled path, and, in this way, identify features such as plot twists, suspenseful moments, etc. Moreover, this representation provides a principled way to obtain the fortune curve described by~\cite{vonnegut1999palm}.

In the game-theoretical representation, the fortune of each character at any given moment in time can be represented very precisely by the expected payoffs under equilibrium starting from that node/info-set, which corresponds to what is called the \emph{value function} in Reinforcement Learning and the \emph{cost-to-go} in control theory. The evaluation of the value function along the tree can also be interpreted as a martingale stochastic process. This perspective allows us to use the methodology by~\cite{ely2015suspense} to measure the `interestingness' of the story in terms of suspense and surprise, as measured in terms of statistical properties of the martingale and the realized sample.

\paragraph{Thought Experiment:} Our long-term vision is to be able to use LLMs to understand incentives faced by real agents and organizations. In a thought experiment, imagine we can process all the emails of a business organization, identify the main decision-makers, the main decisions they need to make, and how they affect each other. Then we may be able to reconstruct which game they are playing and what their incentives are. This could be used to inform leaders about what the hidden incentives are, to advise specific decision-makers about how to play this game, or to design mechanisms to improve outcomes for the organization.

This is, of course, an ambitious, long-term goal. In this paper we take a first step by building a map from stories/text to formal games. Besides informing us about the interaction of decisions and incentives via game-theoretic analysis, such a map from stories to games has other applications as well: It provides a way to classify stories depending on their equilibrium structure (e.g.,~depending on the answers to the questions: Do the characters have a dominant strategy? Does there exist a pure equilibrium or are they forced to randomize? Can their actions be rationalized?). It also allows us to study questions such as how certain game structures lead to more interesting stories.

\paragraph{Roadmap:} In this proof of concept, we use Romeo and Juliet to demonstrate our proposed AI-assisted technique of extracting structure, namely a \emph{game tree}, and then deriving insights from a game-theoretic analysis. Section~\ref{sec:game_tree} explains the process of building out a game tree by prompting Gemini~\citep{geminiteam2024geminifamilyhighlycapable}. Section~\ref{sec:gt_engine} then shows how to collate the information about our game tree and format it for input to a game theory reasoning engine, Gambit~\citep{SavaniTurocy2024}. Section~\ref{sec:results} then interprets the results of the game theory engine's equilibrium computation. We discuss the rationality of the characters as well as conjecture Shakespeare's model of Romeo and Juliet's world based on its consistency with game theory's concept of rationality. Finally, Section~\ref{sec:concl} concludes with a wider discussion of the implications of our proposed approach and avenues for future research.

\section{Game Trees and Nash Equilibrium}

We will formulate a story as a game in extensive form. We refer to the classical textbook of \cite{fudenberg1991game} for a detailed exposition. A game consists of a set of $n$ players (here corresponding to the characters in the story) and is represented as a tree. The root of the tree corresponds to the beginning of the story and each path corresponds to one possible way the story can unfold. The leaf-nodes of the tree correspond to outcomes of the story and we will associate with each outcome a vector of payoffs which can be viewed as how happy each character is with this outcome. The internal nodes of the tree can be of two types: (i) a decision node which corresponds to a decision by one of the players; or (ii) chance node: which is a random decision by nature (capturing e.g.,~whether a secret message sent by one character will reach another character).

In games, it is quite common that not all parties share the same information. For example, in poker, a player does not observe what cards have been dealt to the other players unless that information is revealed in the course of the game. Because of this, as the play unravels from the root of the tree towards a leaf, it is possible that a player cannot distinguish (based on the information they have observed thus far) the true state of the game from several possible nodes based on the information they have observed so far. A set of nodes that would be indistinguishable to a player given the information that they have observed when any one of these nodes would be reached in the course of a play is called an ``information set'' of that player. Each player induces a partition of the nodes of the tree into information sets. Obviously if an information set of some player contains a decision node for that player, then all other nodes in that same information set are also decision nodes for that player and the same actions are available at all decision nodes in the information set. 

A pure strategy of a player in an extensive-form game is a mapping from all information sets containing decision nodes for that player to an action they would take if the play were to reach any one node in that information set (as the player cannot distinguish among these nodes). A mixed strategy of a player is a distribution over pure strategies. A behavioral strategy of a player is a mapping from each information set containing decision nodes for that player to a distribution over the actions that are available to the player at that information set. Once all players choose a strategy, a distribution over leaf nodes is induced, conditioning on the play reaching any particular node of the tree. We can thus compute expected payoffs for all players at each node. We say that a set of strategies are in Nash equilibrium if, given the strategies of the other players, no player can change their own strategy to improve their expected payoff at the root of the tree. In this work, we compute a specific type of Nash equilibrium called a \emph{limiting logit equilibrium} (LLE) that is unique in almost all games~\citep{mckelvey1998quantal,turocy2005dynamic}.

\section{Traversing the Game Tree and Exploring Counterfactuals with an LLM}\label{sec:game_tree}

The first step in our proof of concept is to apply an LLM (in this case Gemini) to identify in a story who are the main characters, what are their decisions, and from there, uncover the plot structure. As a first step we ask Gemini to identify the main characters and their decisions which produces the ``game path'' realized in the play. Next we use the language model to explore how the story could have unfolded if the characters had taken different decisions. The model is asked about what score characters would attribute to each of the outcomes (their game payoffs) as well as the probabilities of certain events (chance nodes). We provide our discussion with Gemini and prompt in the Appendix.

In this proof of concept, there is still a great deal of human guidance in asking the right questions. We take steps to automate this process in Section \ref{sec:concl}. Below we include an excerpt focusing on the point where Juliet decides to move forward with the plan to fake her death (hatched by Friar Lawrence).

{
\begin{itemize}[label={\includegraphics[scale=0.12]{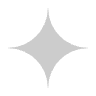}}]\item\begin{displayquote}\small
In Romeo and Juliet, Romeo is banished for avenging Mercutio's death and killing Tybalt. Juliet's family intends for her to marry Paris. Friar Lawrence has other ideas. Name options Juliet could consider including ones that her family might prefer, ones that may appear risky, and ones that might appear tragic to the audience. Please only list three options.
\end{displayquote}\end{itemize}}

Gemini's response:

{\tiny \begin{itemize}[label={\includegraphics[scale=0.12]{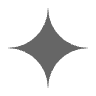}}]\item\begin{displayquote}\small
1. Obey her family and marry Paris (Family's preference): This is the most straightforward and socially acceptable option. It would appease her family and maintain the peace between the Capulets and Montagues. [...]\\\\
2. Fake her own death and reunite with Romeo (Risky): This is Friar Lawrence's plan. It involves Juliet taking a potion that will make her appear dead, allowing her to escape her family and eventually reunite with Romeo in Mantua. While this offers a glimmer of hope for a happy ending, it's incredibly risky. There's a high chance of miscommunication or things going wrong [...].\\\\
3. Take her own life (Tragic): Overwhelmed by the seemingly impossible situation and the prospect of marrying Paris, Juliet might consider suicide as a way to escape her predicament. This would be the most tragic outcome for the audience [...]
\end{displayquote}\end{itemize}}

Gemini indicates several actions Juliet could take. Each action will be modeled as a branch in our game tree. Next we ask Gemini for the probability that her message reaches Romeo in time:

\begin{itemize}[label={\includegraphics[scale=0.12]{figures/gemini/spark_gray.png}}]\item\begin{displayquote} \small
In the case where Juliet chooses to fake her own death, what's the probability that her message alerting Romeo to this plan reaches him in time. Please give a number between 0 and 100.
\end{displayquote}\end{itemize}

Gemini's response:
\begin{itemize}[label={\includegraphics[scale=0.12]{figures/gemini/spark.png}}]\item\begin{displayquote} \small
Given the communication limitations and potential obstacles in Shakespearean times, the probability of Juliet's message reaching Romeo in time is realistically quite low. I would estimate it to be around 30\%.\\\\
Several factors contribute to this low probability: [...]
\end{displayquote}\end{itemize}

Probabilistic events like these are represented with chance nodes; nature rolls the dice and we observe the outcome.

Recall that Juliet is making her initial decision in the aftermath of the traumatic events including deaths of Mercutio, Tybalt, and the banishment of Romeo. We can model the mental impact of this trauma as a probabilistic chance node determining whether Juliet is overwhelmed with grief:

\begin{itemize}[label={\includegraphics[scale=0.12]{figures/gemini/spark_gray.png}}]\item\begin{displayquote}\small
What is the probability that Juliet is initially overwhelmed with grief following the deaths of Mercutio and Tybalt as well as the banishment of Romeo? Assume if she's overwhelmed with grief, she would then decide to take her own life. Please give a number between 0 and 100 for this probability.
\end{displayquote}\end{itemize}

Gemini's response: \includegraphics[scale=0.12]{figures/gemini/spark.png}\textit{I would estimate the probability of Juliet being initially overwhelmed with grief to be around 80-90\%. [...] }.

\section{Creating the Game Object for a Game Theory Engine}\label{sec:gt_engine}

\begin{figure}
    \centering
    \includegraphics[width=0.68\linewidth]{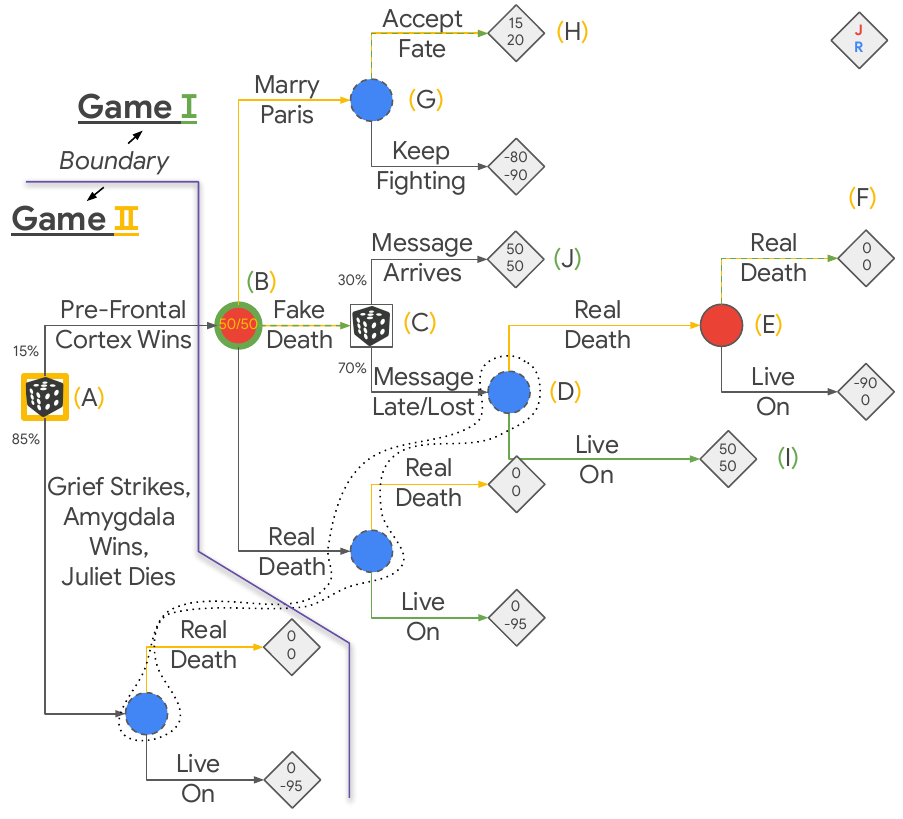}
    \caption{The game trees constructed with the assistance of Gemini for the final plot twists at the end of \textcolor{blue}{Romeo} and \textcolor{red}{Juliet}. We consider both Game~\rom{1} rooted to the right of the purple boundary (the red node (B) highlighted in green) and Game~\rom{2} rooted just left of it (the chance node (A) highlighted in yellow). The Nash equilibrium (specifically, limiting logit equilibrium) of Game~\rom{2} is denoted by the yellow arrows at decision nodes; all strategies at each decision node are deterministic except for Juliet's first decision which flips a fair coin when determining whether to marry Paris or fake her own death. For Game~\rom{1}, the equilibrium strategies are all deterministic and denoted by green arrows. The loop encircling the three blue nodes indicates that these states (histories) appear exactly the same to Romeo; Romeo cannot differentiate between whether or not Juliet has actually died.}
    \label{fig:game_tree:rj}
\end{figure}

Using the information collected with the help of Gemini in the previous section, we can now translate the story into an extensive-form game, which we illustrate in Figure~\ref{fig:game_tree:rj}. The red nodes correspond to decisions by Juliet, the blue nodes are decisions by Romeo and the dice nodes correspond to chance. The grey diamond boxes at the end of each path correspond to the agent's payoffs, i.e., how happy or unhappy each character is with this particular way of the story unfolding.
Notice the dotted loop around the three blue nodes in the diagram. The set of the three nodes is referred to as a single \emph{infostate}, acknowledging that Romeo cannot discern between the distinct histories: has Juliet truly ended her life in grief, or has she merely faked her death without informing Romeo? This unknown is at the heart of Romeo's decision. We examine two models: Game~\rom{2} considers the unpredictable psychological effects of grief on Juliet, whereas Game~\rom{1} assumes perfect agency in actions for Juliet.

\section{Solving the Game and Interpreting Results}\label{sec:results}

\begin{figure}
    \centering
    \begin{subfigure}[t]{0.32\textwidth}
        \centering
        \includegraphics[width=0.95\textwidth]{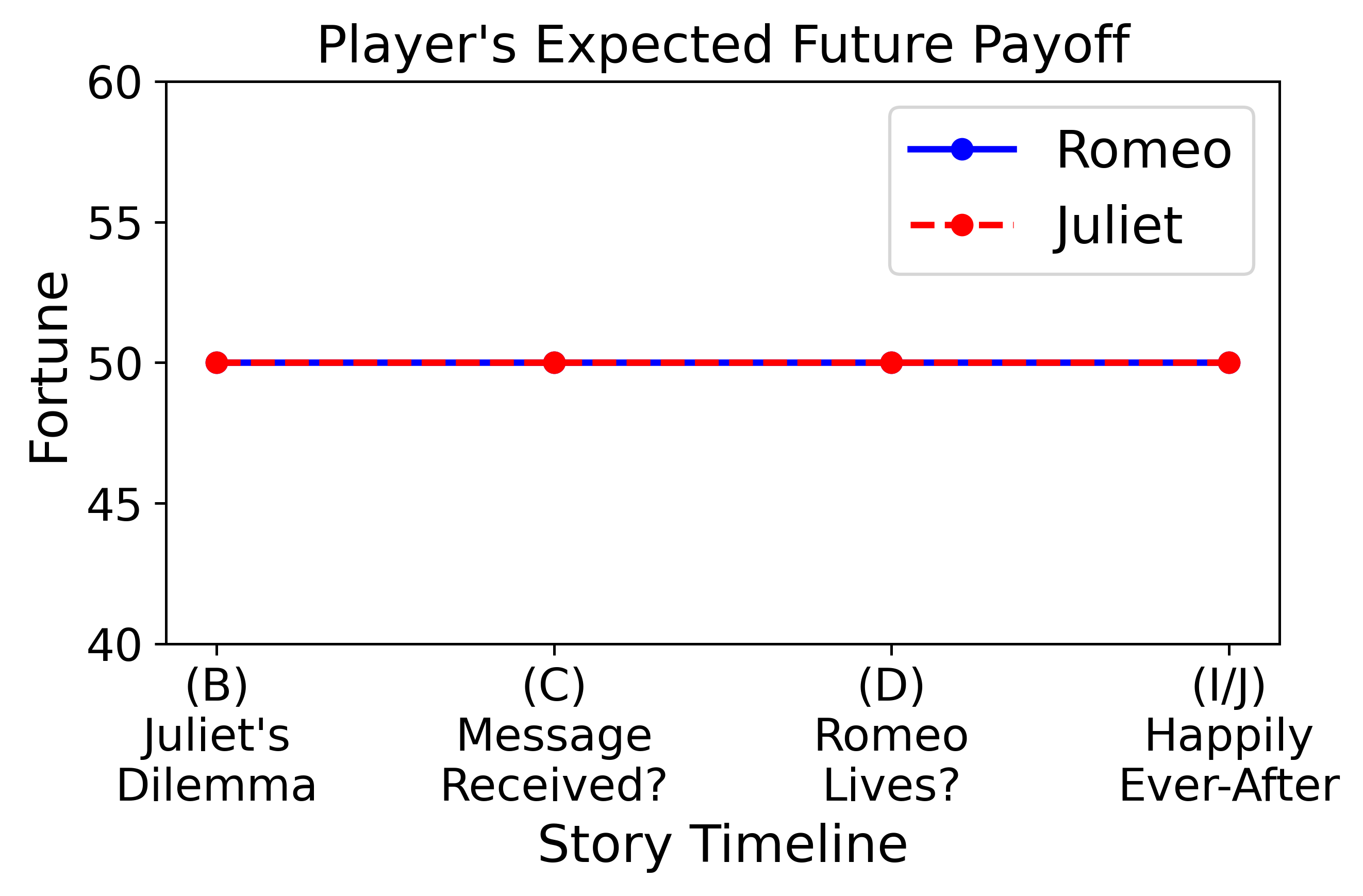}
        \caption{Shape of Game~\rom{1}.}
        \label{fig:shape:small}
    \end{subfigure}%
    ~ 
    \begin{subfigure}[t]{0.32\textwidth}
        \centering
        \includegraphics[width=0.95\textwidth]{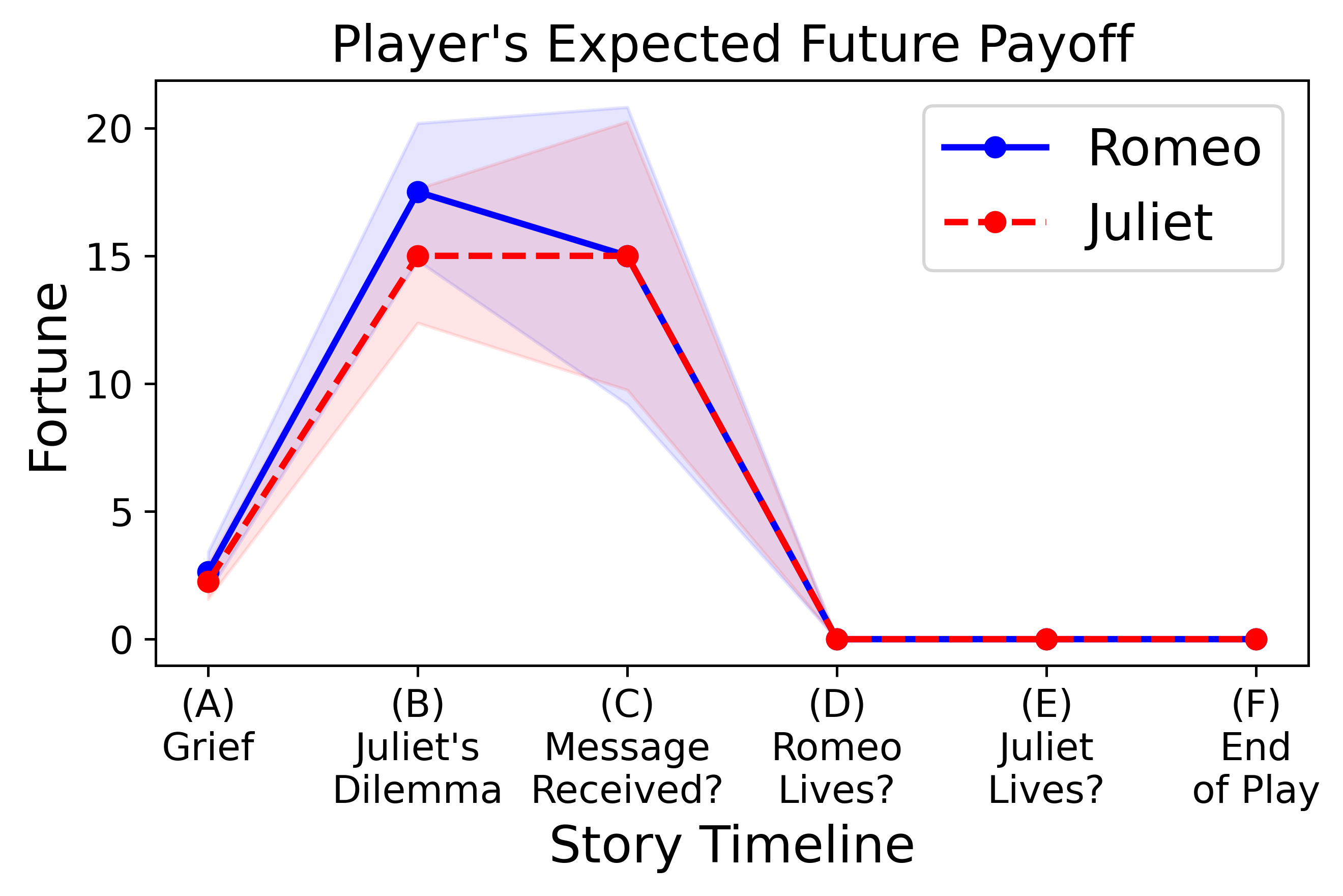}
        \caption{Shape of Game~\rom{2}.}
        \label{fig:shape:big}
    \end{subfigure}
    ~ 
    \begin{subfigure}[t]{0.32\textwidth}
        \centering
        \includegraphics[width=0.95\textwidth]{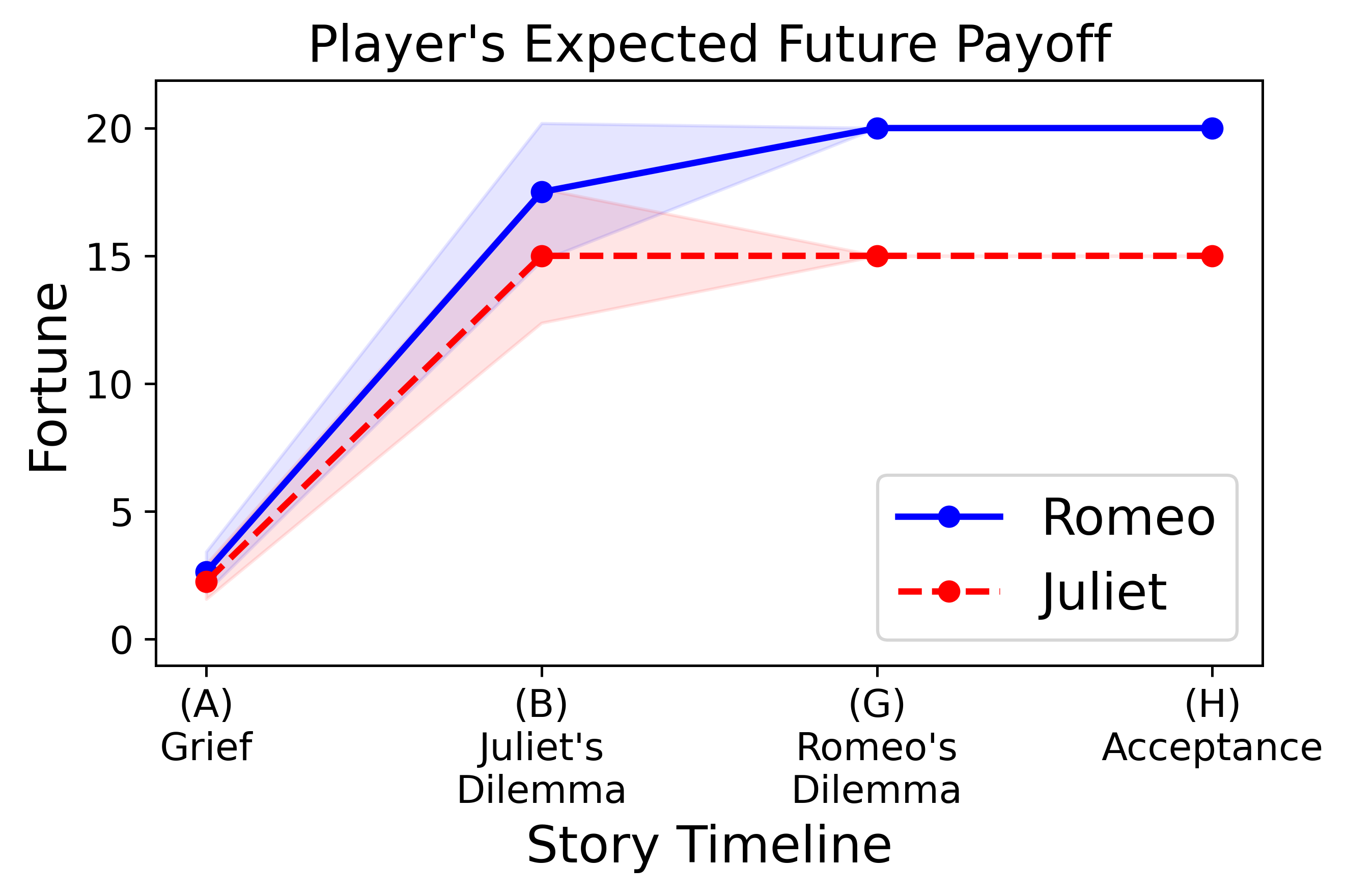}
        \caption{Shape of Game~\rom{2} (Alt).}
        \label{fig:shape:big_b}
    \end{subfigure}
    \caption{\emph{Shape of a Game}: The design of the game tree can have a dramatic impact on the shape of the resulting story. (\subref{fig:shape:small}) Game~\rom{1} has zero surprise (fluctuations in value function) and zero suspense (variance of value function) as it always results in a happy ending (I/J). (\subref{fig:shape:big}) Adding a chance node and decision node to Game~\rom{1} results in Game~\rom{2} and gives the story shape; the martingale of the Romeo and Juliet plot under Game~\rom{2} has both high surprise (large changes in both characters' values after Juliet is overcome with grief at the root node and when the message fails to reach Romeo) and high suspense (high variance of the character's value functions at Juliet's decision to fake her death as well as nature's choice of whether the message successfully reaches Romeo). (\subref{fig:shape:big_b}) Contrast this with the lack of shape of the story given by the alternate ending where Juliet randomly selects to marry Paris. Shakespeare selects the most \emph{interesting} story from the behaviors that are \emph{rationalizable}.}
    \label{fig:shape}
\end{figure}

We solve\footnote{We provide a Google colab for loading and solving games: \url{https://colab.research.google.com/drive/1iPyrjcB95jUZkHfM-cykEdCkE-eQAfz7?usp=sharing}.} the two games using Gambit~\citep{SavaniTurocy2024} and analyze their equilibrium.

In the simpler Game~\rom{1}, we find that the Nash equilibrium (specifically the limiting logit equilibrium) suggests Juliet should fake her death and Romeo choose to live despite his ignorance of her plan.

Contrast this with Game~\rom{2} where Romeo knows there's a chance that Juliet is overcome with grief and chooses to end her misery in death. The Nash equilibrium suggests that with 50\% probability, Juliet elects to fake her death leading to Romeo tragically ending his life. But with equal probability, Juliet elects to marry Paris despite her love for Romeo. According to this game-theoretic model, the audience observes just one possible sequence of events in the story of Romeo and Juliet.

Assuming Game~\rom{1}, Romeo was, in fact, irrational in taking his own life. However, assuming Game~\rom{2}, Romeo was unlucky! Juliet's rational decision was to flip a coin to marry Paris or fake her death, and the coin came up to fake her death. In that case, Romeo's choice to end his misery in death is arguably rational given the payoffs in the game tree representation.

\section{Discussion and Future Impact}\label{sec:concl}

In this proof of concept, we used an LLM to analyze a play, distill its main elements into a game tree and produced two possible game trees (Game~\rom{1} and Game~\rom{2}) to represent the play. Only in Game~\rom{2} are the decisions made by the characters in the actual play in the support of the equilibrium. For that reason, we say that  Game~\rom{2} ``rationalizes'' the story while Game~\rom{1} doesn't. Once we obtain a game tree that rationalizes the story, its equilibrium induces a value function on the 
nodes (Figure \ref{fig:shape}). This allows us to plot how the expected fortune of the characters vary along the game path in the play. Such value function can be thought of as the expectation of a reader (who has the game tree in mind) who is still uncertain at that stage about which path will be taken. In the case of Romeo and Juliet, the value function follows the prototypical path of a tragedy: the fortunes of the characters initially rise, suggesting the possibility of a happy outcome, but ultimately fall precipitously.

Our goal is to develop this proof of concept into a systematic approach that can be readily extended to analyze other stories. This will allow us to obtain game representations for different narratives as well as multiple interpretations of the same story. Analyzing various game theoretic interpretations of a story can reveal different ways to rationalize it.

To facilitate this extension, we can utilize few-shot prompting \citep{Brown2020fewshot} with our Romeo and Juliet game tree example (Section \ref{sec:game_tree}, Figure \ref{fig:game_tree:rj}) and chain of thought prompting \citep{wei2022chain} to guide the model's reasoning about game structure and payoffs. This approach, with light human supervision, has allowed us to generate multiple representations of Romeo and Juliet and extend to new stories like Hamlet (see Appendix and shared colab). 

Several important questions emerge from this work for future investigation. While we defined a game to ``rationalize'' a story when its equilibrium assigns positive probability to the actual story path, evaluating game-theoretic representations requires a more comprehensive framework. Such a framework should consider the structural fidelity to key decision points, the ability to capture strategic interactions between characters, and the validity of assigned payoffs and chance probabilities.

More fundamentally, we acknowledge that game-theoretic representations have inherent limitations in capturing narrative complexity. Elements like character development, emotions, and themes may not readily fit within game trees' formal structure. These boundaries suggest promising directions for developing approaches that complement game-theoretic analysis with other computational methods, potentially offering new insights at the intersection of formal modeling and literary analysis.

\bibliographystyle{abbrvnat}
\bibliography{main}

\newpage
\appendix
\section{Appendix}

Below, we share a prompt used to generate a .efg file capturing a game tree for Hamlet. The prompt includes our original discussion of Romeo and Juliet with Gemini as well as a corresponding .efg file meant as a helpful in-context example to Gemini.

\subsection{Gemini Prompt for Hamlet (Contains Romeo and Juliet Example In-Context)}

\begin{lstlisting}
I want to model the play of Hamlet as a game, specifically I want to construct a .efg file describing a game tree with the characters, their decisions, chance nodes representing the role of nature, etc. I'll first give an example for Romeo and Juliet. Can you then give me a .efg file for Hamlet? Information on the format for the .efg file can be found at https://gambitproject.readthedocs.io/en/v16.2.0/formats.efg.html. Make sure to fill out all terminal outcomes (leaves) of the game tree. Chance nodes (c) contain probabilities but player nodes (p) do not. Nodes in the game tree should be listed in prefix order. Terminal outcomes can be reused and repeated, but they must be listed for every leaf. Put each node on a separate line of the file.

**Discussion used to construct important parts of the game tree for Romeo and Juliet:**

Us:
In Romeo and Juliet, Romeo is banished for avenging Mercutio's death and killing Tybalt. Juliet's family intends for her to marry Paris. Friar Lawrence has other ideas. Name options Juliet could consider including ones that her family might prefer, ones that may appear risky, and ones that might appear tragic to the audience. Please only list three options.

Gemini:
1. Marry Paris (Family's Preference): This option aligns with her family's wishes and societal expectations. It would maintain the peace between the feuding families and secure Juliet's social standing. However, this would betray her love for Romeo and force her into a loveless marriage.

2. Fake Her Own Death (Risky): This is the plan Friar Lawrence devises. By consuming a potion that simulates death, Juliet can escape her forced marriage to Paris and reunite with Romeo. This option is highly risky as it relies on many factors aligning perfectly, including Romeo receiving the message about the plan in time and the potion working as intended.

3. Suicide (Tragic): Overwhelmed by the seemingly insurmountable obstacles to her love for Romeo, Juliet could contemplate taking her own life. This would be the most tragic option, as it would end her life prematurely and cause immense grief for those who love her. This option also goes against the teachings of the church, which would condemn such an act.

Us:
If Juliet marries Paris, what options are available to Romeo? Is love still possible? Please only list two options.

Gemini:
1. Leave Verona and Juliet Behind: Romeo could accept that Juliet has married Paris and move on with his life elsewhere. This would be a painful choice, but it would allow him to avoid further conflict with the Capulets and potentially find love again in the future.

2. Fight for Juliet's Love (Dangerous): Romeo could refuse to give up on Juliet and try to win her back, even though she is married to Paris. This option is incredibly dangerous, as it would likely lead to further violence and potentially result in Romeo's death. However, it is also the only option that could potentially lead to a happy ending for Romeo and Juliet.

Us:
If Romeo thinks Juliet is dead, what famous quotes does Romeo utter and what does it imply about the two options he faces? Please list the two possible actions he can take and the quotes to back them up.

Gemini:
If Romeo believes Juliet is dead, he faces two main options, each reflected in his famous quotes:

1. Suicide:
- Quote: "Well, Juliet, I will lie with thee tonight." This implies that Romeo cannot bear the thought of living without Juliet and chooses to join her in death. He sees no point in continuing his life without his love and decides to end it all.
2. Revenge/Violence:
- Quote: "Then I defy you, stars!" This quote suggests that Romeo blames fate for his misfortune and is ready to challenge it. He may seek revenge against those he perceives as responsible for Juliet's death, leading to further violence and tragedy.
Both quotes highlight Romeo's intense emotions and impulsiveness. His grief over Juliet's supposed death drives him to extreme measures, either to join her in death or to lash out against those he blames for his loss.

Us:
Similarly, if Juliet thinks Romeo is dead, what are the two possible options she faces?

Gemini:
If Juliet believes Romeo is dead, she faces two primary options:

1. Suicide: Overwhelmed by grief and despair, Juliet might choose to take her own life to join Romeo in death. This would mirror Romeo's impulsive reaction and reflect the intensity of their love.

2. Live without Romeo: Juliet could try to continue her life without Romeo, though it would be a life filled with sorrow and regret. She might resign herself to a loveless marriage with Paris or find some other way to carry on. This option would require immense strength and resilience but would allow her to honor Romeo's memory by living a life he would have wanted for her.

Us:
In the case where Juliet chooses to fake her own death, what's the probability that her message alerting Romeo to this plan reaches him in time. Please give a number between 0 and 100.

Gemini:
Given the communication methods and distances involved in Shakespeare's time, the probability of the message reaching Romeo in time is estimated to be around 30%

Several factors contribute to this low probability:

- Distance: Romeo is banished to Mantua, a considerable distance from Verona where Juliet resides.
- Communication method: Messages relied on messengers who could face delays due to travel obstacles, illness, or interception.
- Interference: There's a risk of the message being intercepted or delayed by those who oppose Romeo and Juliet's relationship, such as the Capulets.
However, there's still a chance of the message reaching Romeo in time, as Friar Lawrence's plan relies on this crucial step.

Us:
How much would Romeo value his life going forward if he finds Juliet dead? Please give a number between -100 and 100 with -100 being the lowest value.

Gemini:
-95

Us:
Similarly, how much would Juliet value her life going forward if she finds Romeo dead? Please give a number between -100 and 100 with -100 being the lowest value.

Gemini:
-90

Us:
If Juliet marries Paris and Romeo decides not to immediately fight it, is there a possibility that Romeo would still find love in the future? How much would Romeo value life going forward. Please give a number between -100 and 100 with -100 being the lowest value.

Gemini:
Yes, there is a possibility that Romeo could find love again in the future. While his love for Juliet is intense and passionate, it's not impossible for him to heal and open his heart to someone new. This process might take time and involve significant grief and introspection, but Romeo is young and capable of deep emotion, which could eventually translate into new love.

In this scenario, the value of Romeo's life might be around 20. Though heartbroken and initially devastated, he still has potential for happiness and a future. This value reflects the lingering pain of lost love while acknowledging the possibility of healing and new beginnings.

Us:
What if Juliet married Paris? Could she still have a shot at happiness? Is there a possibility she and Romeo could hatch a plan such that they end up together in the future? Presumably if Romeo can find happiness then so can she, albeit maybe just slightly less than Romeo. Please give a number between -100 and 100 with -100 being the lowest value.

Gemini:
Yes, there's a possibility that Juliet could find happiness even if she marries Paris, although it might not be the passionate love she experienced with Romeo. Here's why:

- Adaptability: Juliet is young and resilient. She might find ways to adapt to her marriage with Paris and potentially develop affection or even love for him over time.
- Secret Communication: Juliet and Romeo could maintain secret communication and potentially plan a future together. While risky, it's not impossible, especially if they have allies like Friar Lawrence.
- Changing Circumstances: Circumstances could change in the future. Paris could die, allowing Juliet to reunite with Romeo, or their families could reconcile, making their relationship more acceptable.

Considering these possibilities, Juliet's happiness level might be around 15. This reflects the initial unhappiness of a forced marriage but acknowledges the potential for adaptation, secret planning, and future changes that could lead to a fulfilling life, even if it's not the one she initially envisioned with Romeo.

Us:
Presumably, Juliet's and Romeo's suicides ends their pain of living in a world without each other. How do you value the absence of pain but also the absence of pleasure? Please give a number between -100 and 100 with -100 being the lowest value.

Gemini:
0

Us:
How pleased would Romeo and Juliet be if they were able to escape and stay together? Would that happiness last forever? What difficulties might they face. What we be a very conservative estimate. Please give a number between -100 and 100 with -100 being the lowest value.

Gemini:
Initially, Romeo and Juliet would be extremely pleased if they were able to escape and stay together, likely at a level of 95. Their love is passionate and all-consuming, and being together would fulfill their deepest desires.

However, this happiness might not last forever. They would face several challenges that could erode their initial joy:

1. Constant Hiding: They would have to live in secrecy, constantly fearing discovery and separation. This could create stress and anxiety, diminishing their happiness over time.
2. Financial Difficulties: Without family support, they would need to find ways to support themselves, which could be difficult and lead to additional stress.
3. Social Isolation: Living in hiding would limit their social interactions and connections, potentially leading to loneliness and isolation.
4. Family Disapproval: The ongoing feud between their families would remain a constant threat, and they would likely never be fully accepted by either side.

Considering these challenges, a conservative estimate of their long-term happiness would be around 50. While they would be together, the difficulties they face would significantly impact their overall happiness and well-being.

Us:
Alright. What would happen if Romeo refused to accept Juliet's marriage and disrupted the wedding?

Gemini:
If Romeo disrupted Juliet's wedding to Paris, it would likely lead to a chaotic and violent confrontation. Here's how the scenario might unfold:

1. Chaos and Interruption: Romeo's sudden appearance would shock everyone present, especially Juliet and her family. The wedding ceremony would be abruptly interrupted, causing confusion and alarm among the guests.

2. Confrontation with Paris: Romeo and Paris, both claiming Juliet as their own, would likely engage in a heated argument or even a physical fight. This confrontation could escalate quickly, potentially leading to injuries or even death.

3. Escalation of the Feud: Romeo's actions would further enrage the Capulets, who would see this as a deliberate insult and a threat to their family's honor. This could reignite the feud with the Montagues, leading to further violence and bloodshed.

4. Legal Consequences: Romeo's disruption of a lawful ceremony and potential violence against Paris could lead to severe legal consequences. He could face arrest, imprisonment, or even execution for his actions.

5. Juliet's Dilemma: Juliet would be caught in a difficult situation, torn between her love for Romeo and her obligations to her family and new husband. Her choices in this moment could have significant consequences for her own future and happiness.

Overall, Romeo's disruption of Juliet's wedding would have disastrous consequences for everyone involved. It would escalate the existing tensions between the families, lead to violence and potential deaths, and put Romeo and Juliet's future in jeopardy.

Us:
How would Juliet fair in this outcome? Please give a number between -100 and 100 with -100 being the lowest value.

Gemini:
Juliet's happiness in this outcome would be extremely low, likely around -80. Here's why:

- Public Humiliation: The disruption of her wedding would be a public spectacle, causing her immense shame and embarrassment.
- Family's Wrath: Her family would be furious with her for Romeo's actions, potentially leading to harsh punishment or even disownment.
- Lost Reputation: Her reputation would be tarnished, making it difficult for her to find a suitable match in the future.
- Guilt and Grief: If Romeo is harmed or killed in the ensuing violence, Juliet would feel immense guilt and grief, further deepening her despair.
- Uncertain Future: Her future would be uncertain and bleak, with limited options and little hope for happiness.

Overall, this outcome would be devastating for Juliet, leaving her isolated, humiliated, and heartbroken.

Us:
How would Romeo fair?

Gemini:
Romeo's fate in this outcome would be equally bleak, likely around -90. Here's why:

- Physical Harm or Death: Engaging in a confrontation with Paris and potentially other Capulets could lead to serious injury or death.
- Legal Consequences: His actions would result in severe legal repercussions, including imprisonment or execution.
- Loss of Juliet: Even if he survives, his relationship with Juliet would be irrevocably damaged. Her family would never accept him, and their love would be shrouded in secrecy and danger.
- Guilt and Regret: If his actions lead to Juliet's suffering or harm, he would be consumed by guilt and regret.
- Isolation and Desperation: He would be forced to live in hiding, constantly fearing capture and separated from his loved ones.

Overall, this outcome would be disastrous for Romeo, leading to a life of isolation, regret, and potential violence.

Us:
What is the probability that Juliet is initially overwhelmed with grief following the deaths of Mercutio and Tybalt as well as the banishment of Romeo? Assume if she's overwhelmed with grief, she would then decide to take her own life. Please give a number between 0 and 100 for this probability.

85

**Corresponding .efg File**:

EFG 2 R "Romeo and Juliet" { "Juliet" "Romeo" }
"Romeo and Juliet extensive form game"

c "" 1 "Grief" { "Rational" 15/100 "Suicide" 85/100 } 0

p "Start" 1 2 "Start Info State" { "Marry Paris" "Fake Death" "Suicide" } 0
p "" 2 3 "Marry Paris" { "Accept Fate" "Fight" } 0
t "" 1 "Outcome 1" { 15, 20 }
t "" 2 "Outcome 2" { -80, -90 }

c "" 4 "Fake Death" { "Message Received" 30/100 "Message Not Received" 70/100 } 0

t "" 3 "Outcome 3" { 50, 50 }

p "" 2 5 "Thought Juliet Dead" { "Suicide" "Live" } 0
p "Juliet Decides" 1 6 "Message Not Received, Romeo committed suicide" { "Suicide" "Live" } 0
t "" 4 "Outcome 4" { 0, 0 }
t "" 5 "Outcome 5" { -90, 0 }
t "" 3 "Outcome 3" { 50, 50 }

p "" 2 5 "Thought Juliet Dead" { "Suicide" "Live" } 0
t "" 4 "Outcome 4" { 0, 0 }
t "" 6 "Outcome 6" { 0, -95 }

p "" 2 5 "Thought Juliet Dead" { "Suicide" "Live" } 0
t "" 4 "Outcome 4" { 0, 0 }
t "" 6 "Outcome 6" { 0, -95 }
\end{lstlisting}

\end{document}